\documentclass[fleqn,10pt]{wlscirep}
\usepackage[utf8]{inputenc}
\usepackage[T1]{fontenc}
\usepackage{mathtools}
\usepackage{times}
\usepackage{lineno}
\usepackage{xcolor}
\usepackage{xspace}
\usepackage{hyperref}
\usepackage{comment}
\usepackage{soul}

\newcommand{\ie}{\emph{i.e.}}
\newcommand{\eg}{\emph{e.g.}}
\newcommand{\ea}{\emph{et al.}}

\newcommand{\ER}{Erd\"{o}s-R\'enyi }
\newcommand{\avg}[1]{\langle #1\rangle}

\title{Self-induced consensus of Reddit users to characterise the GameStop short squeeze}

\author[1,2,+]{Anna Mancini}
\author[1,2,+]{Antonio Desiderio}
\author[3,4,*]{Riccardo Di Clemente}
\author[1,2,$\ddag$]{Giulio Cimini}
\affil[1]{Physics Department and INFN, Unversità di Roma Tor Vergata, 00133 Rome (Italy)}
\affil[2]{Centro Ricerche Enrico Fermi, 00184 Rome (Italy)}
\affil[3]{Department of Computer Science, University of Exeter, (United Kingdom)}
\affil[4]{The Alan Turing Institute, London NW12DB (United Kingdom)}
\affil[*]{Corresponding Author: r.di-clemente@exeter.ac.uk}
\affil[$\ddag$]{Corresponding Author: giulio.cimini@roma2.infn.it}
\affil[+]{these authors contributed equally to this work}

\date{}

\begin{abstract}
The short squeeze of GameStop (GME) shares in mid-January 2021 has been primarily orchestrated by retail investors of the Reddit r/wallstreetbets community. As such, it represents a paramount example of collective coordination action on social media, resulting in large-scale consensus formation and significant market impact.
In this work we characterise the structure and time evolution of Reddit conversation data, showing that the occurrence and sentiment of GME-related comments (representing how much users are engaged with GME) increased significantly much before the short squeeze actually took place. 
Taking inspiration from these early warnings as well as evidence from previous literature, we introduce a model of opinion dynamics where user engagement can trigger a self-reinforcing mechanism leading to the emergence of consensus, which in this particular case is associated to the success of the short squeeze operation. 
Analytical solutions and model simulations on interaction networks of Reddit users feature a phase transition from heterogeneous to homogeneous opinions as engagement grows, which we qualitatively compare to the sudden hike of GME stock price. 
Although the model cannot be validated with available data, it offers a possible and minimal interpretation for the increasingly important phenomenon of self-organized collective actions taking place on social networks.
\end{abstract}

\begin{document}

\setstcolor{red}

\flushbottom
\maketitle
\thispagestyle{empty}

\section*{Introduction}

Online social media have revolutionized the way we interact with peers, share information and form opinions \cite{kaplan2010users,heidman2012online}, giving rise to new large-scale social phenomena: for instance, the spreading of fake news \cite{delvicario2016spreading,lazer2018science}, the formation of echo chambers and polarized opinions \cite{bakshy2015exposure,bail2018exposure}, the organization of collective actions --- from the Arab Spring \cite{wolfsfeld2013social} to climate change protests \cite{segerberg2011social}. 
Recently an online mass coordination on {\small \href{https://reddit.com/r/wallstreetbets}{\texttt{r/wallstreetbets}}} (WSB), a community of the social media platform Reddit, was able to trigger a short squeeze of GameStop shares with a large impact on financial markets \cite{gamestop_wiki}. 

Reddit is a public discussion website whose users interact with each other by submitting new \emph{posts} and adding \emph{comments} to existing posts or comments, thus creating tree-structured conversation threads. 
The forum is organised into various independent \emph{subreddits}, each dedicated to a specific topic. The subreddit WSB is a community where users (retail investors but also non-skilled traders who use no-commission mobile apps such as \url{robinhood.com}) discuss 
trading strategies and share their gains and losses. The hallmarks of WSB are the irreverent jargon and edgy humor used in conversations \cite{boylston2021wallstreetbets}, as well as the gambling attitude of its users who yet seem to give good investment advice \cite{buz2021investment}.
The popularity of this forum has steadily increased in recent years and has exploded after the events of the GameStop saga.

GameStop (NYSE:GME) is a U.S. video game retailer which was struggling in recent years due to competition from digital distribution services as well as economic effects of the COVID-19 pandemic. As a result GME stock price declined (reaching an all-time low of \$2.57 on April 3, 2020), leading many hedge funds to short sell the stock --- meaning they would profit from its further decrease in price. On the contrary WSB users, likely driven by the opportunity to make profit and possibly anger towards institutional investors \cite{chohan2021counter}, coordinated with the intent to trigger a short squeeze, \ie, a rapid increase in the stock price due to the excess of demand and lack of availability. 
The resulting large-scale mass coordination (buying and holding GME shares) succeeded in driving up the price of GME, attracting even more users and forcing short sellers to cover their positions at large losses, thus further promoting the price rally. On January 28, 2021 GME shares reached an astounding high price of \$483.00; more than 1 million of its shares were deemed failed-to-deliver, which sealed the success of the short squeeze.

Such a highly coordinated financial `operation' received a huge attention not only from the media and financial stakeholders, but from the academic community as well. Following a popular stream of literature aimed at predicting stock market trends using social network moods \cite{bollen2011twitter,broadstock2019social} (with a recent focus on WSB and price movements of cryptocurrencies \cite{phillips2017predicting,wooley2019extracting,lamorgia2021doge}), most of the scholars' attention has been devoted to understanding whether WSB activity, conversation sentiment and user interactions could be used to predict retail trading activity and GME returns, using linear regression models or machine-learning approaches \cite{lyocsa2021YOLO,anand2021wallstreet,betzer2021if,hu2021rise,long2021just,wang2021predicting,semenova2021reddits}. 
Only a few empirical works address the more fundamental question of how coordination or consensus could spontaneously emerge in this context. 
Boylston \ea \cite{boylston2021wallstreetbets} show that the jargon and humor of WSB members is a way to express and reinforce the community's sense of identity. Semenova and Winkler \cite{semenova2021reddits} find empirical evidence of psychological contagion among WSB users, where an initial set of investors attracts a larger and larger group of excited followers --- net of any fundamental price movements. 
Lucchini \ea \cite{lucchini2021from} measure the growing user commitment to the GME operation and social identity of WSB participants, describing the key role of a committed minority of users in triggering the collective action. Anand and Patak \cite{anand2021wallstreet} go along the same line showing that it was a tiny minority of 462 most influential subredditors whose posts most impacted the GME stock price. 

In this paper we provide new empirical evidence on how consensus on the GME operation emerged in the WSB community. We analyse discussions on WSB from September 01, 2019 to February 01, 2021 (see Methods), characterising how the forest of tree-like conversation threads grew as the GME saga unfolded. 
We measure user engagement towards GME through the occurrence and mean sentiment of GME-related conversations. These variables increased significantly far before January (in particular, the frequency of GME in conversations peaks in correspondence to the major events in the GameStop saga) and thus provide early signs of the collective action.

All these empirical evidences suggest that an endogenous and self-reinforcing effect played a fundamental role in triggering consensus formation on the GME operation. 
A possible way to include this mechanism in classical models of opinion dynamics \cite{lazer2009computational,castellano2009statistical,acemoglu2011opinion,noorazar2020recent,baronchelli2021emergence} is through a self-induced global field that drives users towards collective unity \cite{kearns2009behavioral}. 
However, existing models that couple peer interaction to community-wide effects typically consider the presence of external fields acting in the same way on all (or groups of) users, describing the effect of conventional media or other exogenous factors \cite{michard2005theory,bhat2019nonuniversal,majmudar2020voter}. Other models use endogenously-generated fields that align the user opinion with the one held more frequently in the past, in order to mimic the effect of recommendation systems on opinion formation \cite{demarzo2020emergence}. Instead, in order to model opinion formation within the GameStop saga, we assume that a high and widespread engagement with the GME collective operation increases the likelihood that users themselves become committed and will actively participate to the short squeeze --- since its success strongly depends on the number of participants.

Therefore we propose to model opinion dynamics in this scenario considering that users form their opinions either by interacting with peers or by following a global field, which is self-induced by the current status of the community and whose strength is determined by the mean level of user engagement.
Analytical mean-field solutions of the model display a phase transition from a disordered state (where no opinion prevails) to full consensus as user engagement grows. 
Model simulations on statistically validated social networks of WSB users, extracted from their `reply-to' interaction patterns, feature a broader transition that implies a non negligible level of consensus even when engagement is low. 
However the transition becomes abrupt when, as data suggests, the community grows together with the level of consensus reached.

Unfortunately, to fully validate the model we would need access to confidential information on actual purchases of GME shares by WSB users that actual laws prohibits to collect. Therefore the model stands as a minimal yet solvable framework that offers only a possible way to qualitatively reproduce the explosive dynamics of the GameStop event. 
Nevertheless the assumption of a self-induced field, which is supported by the empirical analysis, leads to a spontaneous phase transition from disorder to order, which is seldom found in models of opinion dynamics.

\begin{figure}[ht!]
    \centering
    \includegraphics[width=\textwidth]{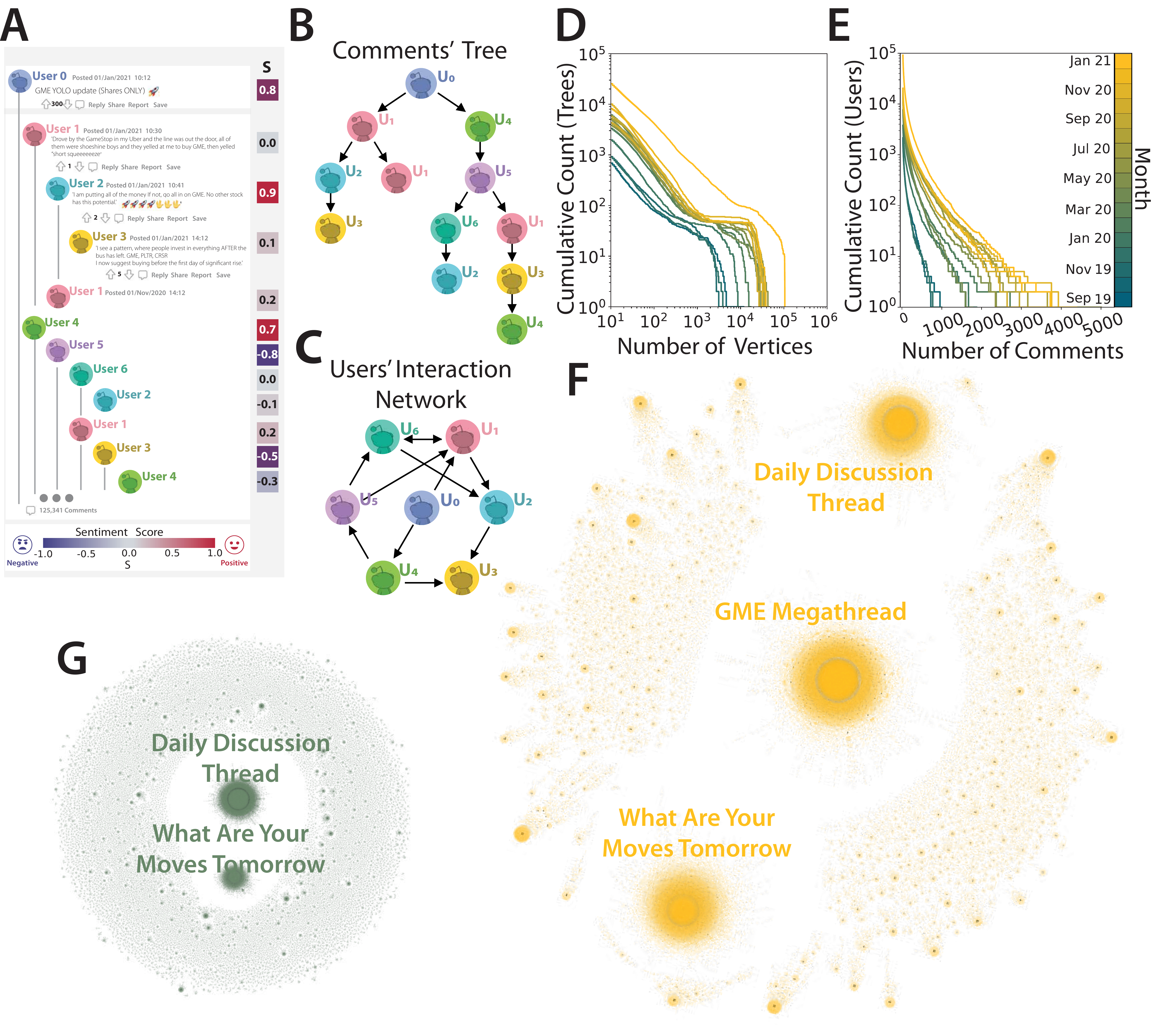}
    \caption{\textbf{Tree structure of Reddit conversations.} 
A) Reddit users can write posts to initiate a conversation and can comment (reply) to posts or other comments. A sentiment score between -1 and +1 (negative and positive, respectively) can be computed for each post and comment. B) The conversation on the social network can be represented by a forest of {\em trees}, where each post is the root of a tree and subsequent comments represent its branches. C) Trees can be used to extract a user-user network, where a link from user $i$ to user $j$ represents the number of times $i$ commented to posts/comments by user $j$. D) Cumulative count of trees larger than a given size (in terms of number of comments) for different months in the considered time span of data. E) Cumulative count of users that contributed more than a given number of posts/comments during different months. F, G) Sample representation of the forest of tree conversations for January 21, 2021 and March 19, 2020, respectively. Icons in figure A-B-C have been designed by Freepik. The schematic representation of a reddit post in figure A has been recreated by the authors. The network representations on figure F-G have been created by the authors using the software Gephi.}
    \label{fig1}
\end{figure}

\section*{Results}

\subsection*{Reddit conversation patterns.}

Figure \ref{fig1}A shows the typical structure of a Reddit post with the comment section underneath. Figure \ref{fig1}B highlights how this structure can be translated into a forest of trees: each post corresponds to the root of a tree, while comments to this post or to other comments in the same thread represent the tree branches. 
Figure \ref{fig1}C shows how these trees can be used to extract a network of user-user 'reply to' interactions (which we shall discuss later on). 
Visual inspection of daily forests, each containing all trees rooted in posts published on the given date, gives a first idea of how the structure of WBS conversations looks like and how it has evolved over time. 
Figure \ref{fig1}F shows the forest of Jan 21, 2021 --- the day before the short squeeze was initiated, while Figure \ref{fig1}G shows the forest of March 19, 2020, when WSB was not as popular and its activity much less intense. We see how the daily forest has grown substantially in terms of overall number of comments as well as number and size of trees. 
Note in particular how each daily forest is characterized by a giant tree: the \emph{Daily Discussion Thread}, created with the purpose of summarizing the events of the day and planning future actions \cite{boylston2021wallstreetbets}, where users are encouraged to comment by WSB rules. Other very large trees are often present, such as \emph{What Are Your Moves Tomorrow}, whereas the \emph{GME Mega-thread} appears in the daily discussions of January.
Figure \ref{fig1}D shows the monthly histograms of the cumulative number of trees by size, whose power law trends end at large sizes due to deviations produced by such mega-threads. 
Moreover, the overall number of conversation trees grows in time. The reason is not only an increasing number of WSB users who join the discussion, but also their increasing activity in terms of number of contributed posts or comments. This can be seen in the monthly histograms of the 
cumulative number of users by number of contributions shown in Figure \ref{fig1}E.
In this plot, the area under the curves that grows in time indicates an increasing number of active users, whereas the increasingly fat tail signals that these users are also contributing with more posts and comments.
Further analyses on conversation trees are reported in the Supplementary Note 1.

\begin{figure}[ht!]
    \centering
    \includegraphics[width=\textwidth]{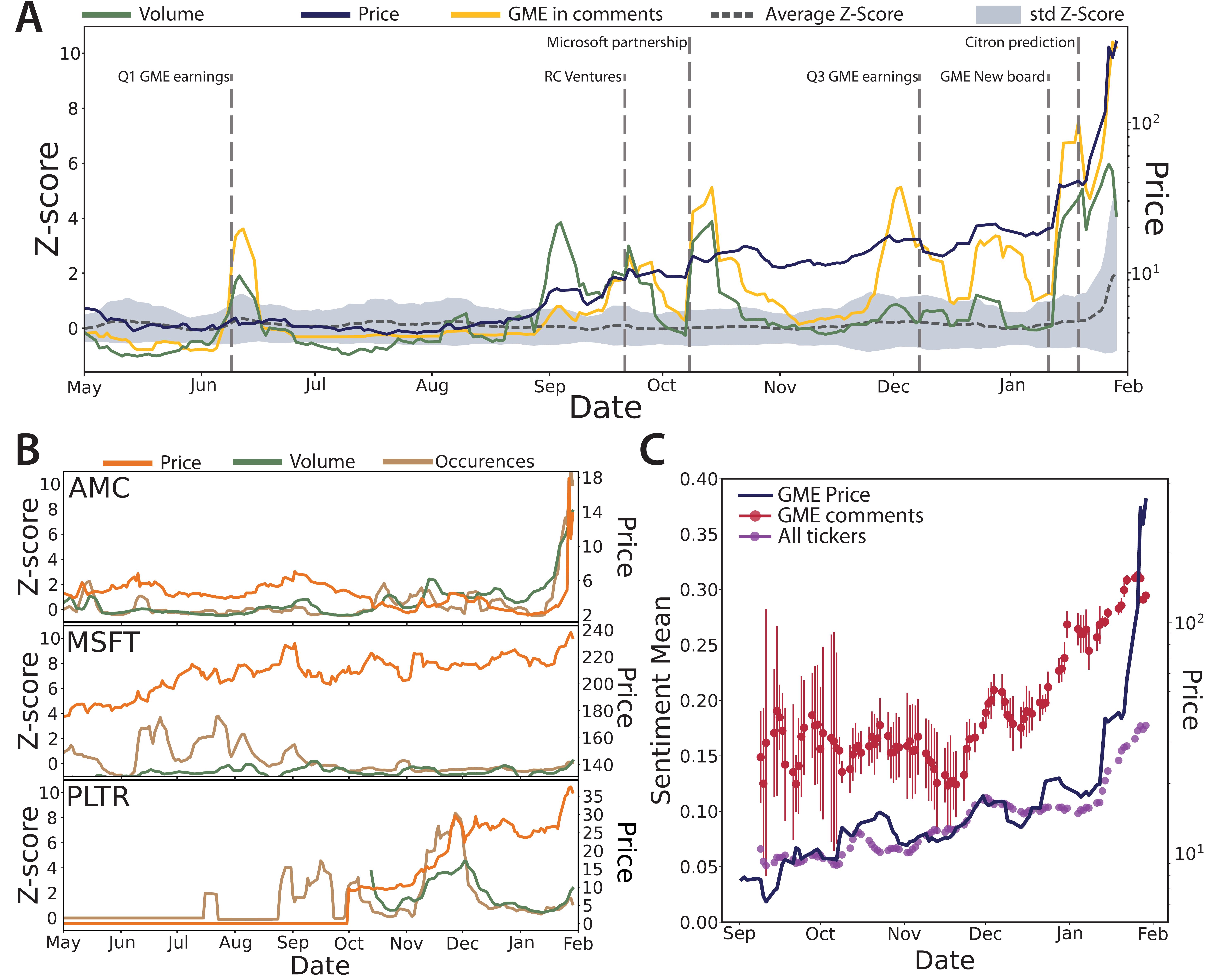}
    \caption{\textbf{Content and sentiment of WSB conversations across the GME saga.} In the following plots we do not consider daily data for weekends and US holidays, when activity on WSB is lower as the stock exchange is closed. We also apply a 5-day moving average to the time series. A) Z-score for the occurrences of `GME' in WSB conversations, compared to the mean Z-score for the occurrence of all stock tickers (shaded area). GME peaks correspond to major events in the GameStop saga. We contextualize these events by also reporting the Z-score of GME trading volumes and the closing price of GME shares.
    B) For other three representative stocks (AMC, MSFT, PLTR): Z-score of the ticker occurrences in conversations, Z-score of trading volumes and closing price. 
    C) Mean sentiment (and standard deviation of the mean) of comments containing `GME', with respect to the same quantity computed on all comments. GME closing price is also reported for illustrative purposes.}
    \label{fig2}
\end{figure}

\subsection*{Conversation content and sentiment.} 
We now turn to the analysis of the content of WSB conversations and how it evolves in time. 
Since WSB is a community of users focused on trading, the occurrence of stock tickers in the text of posts and comments represents a first indicator of what is a popular conversation topic. The total number of occurrences for the various stock tickers follows a power law (Supplementary Note 2), with some very large outliers --- GME in particular is the most frequent one. In order to detect statistically significant occurrences in time we compute their daily Z-scores (see Methods). Figure \ref{fig2}A shows the Z-scores for GME, compared to the average Z-score of all tickers \cite{toole2012inferring}. We clearly see how the peaks given by significant Z-scores correspond to major events of the GameStop saga: (2020-06-09) GME Q1 earning reports; (2020-09-21) RC Ventures increases its stake in GME to 9.98\%; (2020-10-08) GameStop announces a multiyear strategic partnership with Microsoft; (2020-12-08) GME Q3 earning reports, with $257\%$ increase in e-commerce revenues; (2021-01-11) GME announced a new Board of Directors; (2021-01-19) Citron Research predicted that GME's price would fall and belittled GME buyers on Twitter \cite{citron_gme}. 
Notably these peaks become higher in time, signaling that the community's interest towards GameStop grows substantially until January, when GME monopolizes the conversation on WSB. 
Additionally these peaks mostly coincide with those for the Z-score of GME trading volume (\ie, the number of shares traded daily), pointing to a strong relation between the two variables. 
A similar but weaker signal can be found regarding conversations about other stocks, as shown in the examples reported in Figure \ref{fig2}B. AMC (AMC Entertainment Holdings Inc) is a penny stock that similarly to GME was suffering due to the COVID-19 pandemic and was then subject to a short squeeze in mid February 2021 \cite{lyocsa2021YOLO}.
This event is not covered by our data, but we can already see a significant signal of AMC occurrences at the end of January. 
MSFT (Microsoft Corporation) is instead a more solid stock with a constant and regular price growth; in this case we do not observe significant occurrences. 
At last PLTR (Palantir Technologies Inc) had its public debut at the end of September 2020, yet it is remarkable that a significant interest from WSB users was present in the previous months. The peak in the second half of November 2020 was due to a new contract of the company with the U.S. Army and a consequent price jump of +170\%, after which Citron Research labeled the stock as a gambling deal \cite{citron_palantir}.

Besides assessing the content of posts/comments by WSB users we also look at their \emph{sentiment}. As discussed in the introduction, this variable has been often pointed out as a predictor of market movements. We thus perform text sentiment analysis using VADER (\emph{Valence Aware Dictionary and sEntiment Reasoner}) \cite{hutto2014VADER}, a python tool that assigns to each piece of text a score between -1 (very negative) and +1 (very positive). In line with other studies \cite{wang2021predicting,anand2021wallstreet}, we adapt the VADER dictionary to the peculiar jargon and sarcasm used by WSB members (see Methods and Supplementary Note 3). Figure \ref{fig2}C shows an intensive sentiment indicator, \ie, the mean sentiment of all daily posts/comments that mention GME. We see that the signal is initially quite noisy due to the low number of GME-related comments until mid-October; Then as early as the beginning of December it starts to grow significantly (both with respect to its previous trend and to the mean sentiment of all comments), far before the short squeeze of January. Overall we can associate these empirical evidences to a growing engagement of users with GME, which in turn represents an early sign of consensus formation in the community concerning the short squeeze operation.
In light of these results, we now work out a model in which user engagement with a collective cause can influence opinion dynamics and foster the emergence of consensus (or cooperation) thanks to a self-induced feedback mechanism.

\begin{figure}[ht!]
    \centering
    \includegraphics[width=\textwidth]{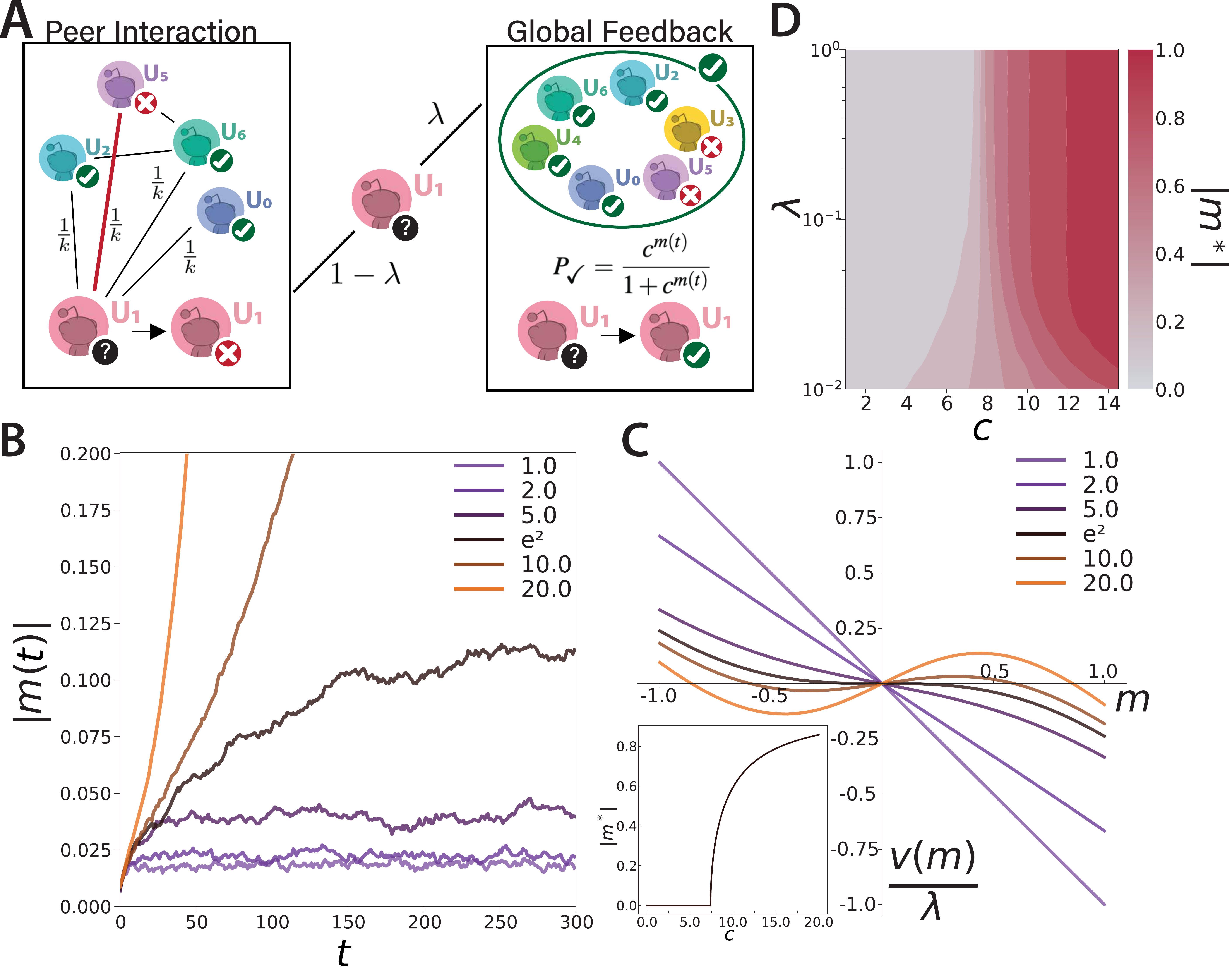}
    \caption{\textbf{Voter model with self-induced global feedback.}
A) Schematic representation of the update rule of the model. At each time step, a user takes on the opinion of either a randomly chosen neighbor (with probability $1-\lambda$) or is influenced (according to a control parameter $c$) by the current level of consensus in the community, namely the magnetization $m$. B) Sample stochastic realizations of the model dynamics: temporal evolution of the magnetization $m$ for different values of the parameter $c$ setting the strength of the global feedback, for $\lambda=0.1$. C) Re-scaled drift term $v(m)$ of the model dynamics (according to the mean-field approximation, for $\lambda\neq0$) as a function of the magnetization of the system. Inset: stable equilibrium points $|m^*|$ as a function of $c$. D) Phase diagram of the model simulated on \ER random graphs of $N=10000$ nodes and average degree $\avg{k}=20$. Icons in figure A have been designed by Freepik.}
    \label{fig3}
\end{figure}

\subsection*{Voter model with self-induced global feedback.} 
We build on one of the most popular theoretical frameworks of opinion dynamics: the \emph{voter model} \cite{clifford1973model,holley1975ergodic}. In the standard voter dynamics, $N$ users are placed on the nodes of a network and are endowed with a binary opinion $s\in\{-1,+1\}$. Starting at $t=0$ from an initially disordered configuration where each user $i$ has opinion $s_i(0) = \pm 1$ with equal probability, at each time step $\delta t = \frac{1}{N}$ a user is chosen at random and copies the opinion of one of her neighbors. The \emph{magnetization} or \emph{order parameter} $m(t)=\frac{1}{N}\sum_is_i(t)$ represents the average opinion at time $t$, or equivalently the level of consensus reached, with $m(t)\simeq0$ and $m(t)=\pm 1$ indicating no consensus and full consensus, respectively. 
The standard model has been studied extensively on different population structures and has been adapted to a variety of different situations \cite{castellano2009statistical,redner2019reality}.
We are interested in a model formulation where, depending on how much users are engaged with a collective cause, they are more keen on assuming a given opinion if that opinion is popular within the community. Mathematically speaking, the model should include a tunable self-induced field acting on all users simultaneously. 
Following an approach formally similar to \cite{demarzo2020emergence}, we define the update rule as: 
\begin{equation}\label{eq:updaterule}
s_i(t + \delta t) = 
    \begin{cases}
        s_j(t) & \; \text{with probability} \; \frac{1 - \lambda}{k_i}\\
        e(t) & \; \text{with probability} \; \lambda
    \end{cases}
\end{equation}
This expression, visually represented in Figure \ref{fig3}A, has the following meaning. 
When user $i$ is selected for the update, with probability $1 - \lambda$ she copies the state of a random neighbor $j$ (\ie, each neighbor is selected with probability $\frac{1}{k_i}$, where $k_i$ is the \emph{degree} or number of neighbors of $i$). 
Instead, with probability $\lambda$ she follows a global field given by the random variable $e(t)\pm 1$. In order to have a self-induced field depending on the current level of consensus, we impose that the probability of $e(t)=+1$ is
\begin{equation}\label{eq:autofield}
    P_1[e(t)] = \frac{c^{m(t)}}{1 + c^{m(t)}}
\end{equation}
When there is no consensus at all (\ie, $m(t)=0$) we have $P_1[e(t)]=\frac{1}{2}$: the global field acts randomly on each user and is equivalent to a white noise term. Instead $m(t)\to+1$ leads to $P_1[e(t)]\to 1$ and analogously $m(t)\to-1$ to $P_1[e(t)]\to 0$: the global field is increasingly able to align users with the majority opinion. $c\geq 1$ is a control parameter that we associate with the level of user engagement: the higher the value of $c$, the less consensus is required for users to align with $m(t)$.

We can understand the behavior of the model for different values of $c$ through its analytical mean-field solution (see Methods). Figure \ref{fig3}B shows sample realizations of the stochastic temporal dynamics of the magnetization (for a fixed value of $\lambda=0.1$), while Figure \ref{fig3}C shows the drift term $v(m)$ rescaled by $\lambda>0$ as a function of $m$. 
For $c=1$ we have $P_1[e(t)]=\frac{1}{2}$: the global field is always white noise that keeps the system in the initial disordered configuration, as in the \emph{noisy voter model} \cite{kirman1993ants,granovsky1995noisy}. 
This is due to the drift and magnetization having always opposite sign: the process is mean-reverting and the only equilibrium point is $m^*=0$. 
Such equilibrium remains stable also when $c>1$, though the drift towards it becomes less intense. 
At the singular point $c=e^2$ we have $P_1[e(t)] = \frac{1}{2}[1+\tanh\,m(t)]$ and the drift vanishes in the region around $m=0$: the initial stochastic dynamics becomes purely diffusive, as in the standard voter model \cite{castellano2009statistical}. 
Finally for $c$ above this threshold the drift pushes the system away from $m=0$ with a speed that grows with $\lambda$: the dynamics quickly reaches a new stable equilibrium point that becomes closer to full consensus as $c$ grows. 
Looking at the stable states $|m^*|$ of the dynamics as a function of $c$ (inset of Figure \ref{fig3}C) we see that the system exhibits a classic second order phase transition from disorder to order.
These results are confirmed by numerical simulations of the model on \ER random graphs (Figure \ref{fig3}D). Only for very small values of $\lambda<0.1$, for which the interaction between peers is largely dominant, network effects make the transition less sharp. 
Furthermore, in this region the dynamics is very slow so the system keeps memory of its initial configuration for a long time. 
This produces finite-time hysteresis loops that slow down both the emergence of consensus from a disordered configuration and its dissolution from an ordered one (see Supplementary Note 4).

\begin{figure}[ht!]
    \centering
    \includegraphics[width=\textwidth]{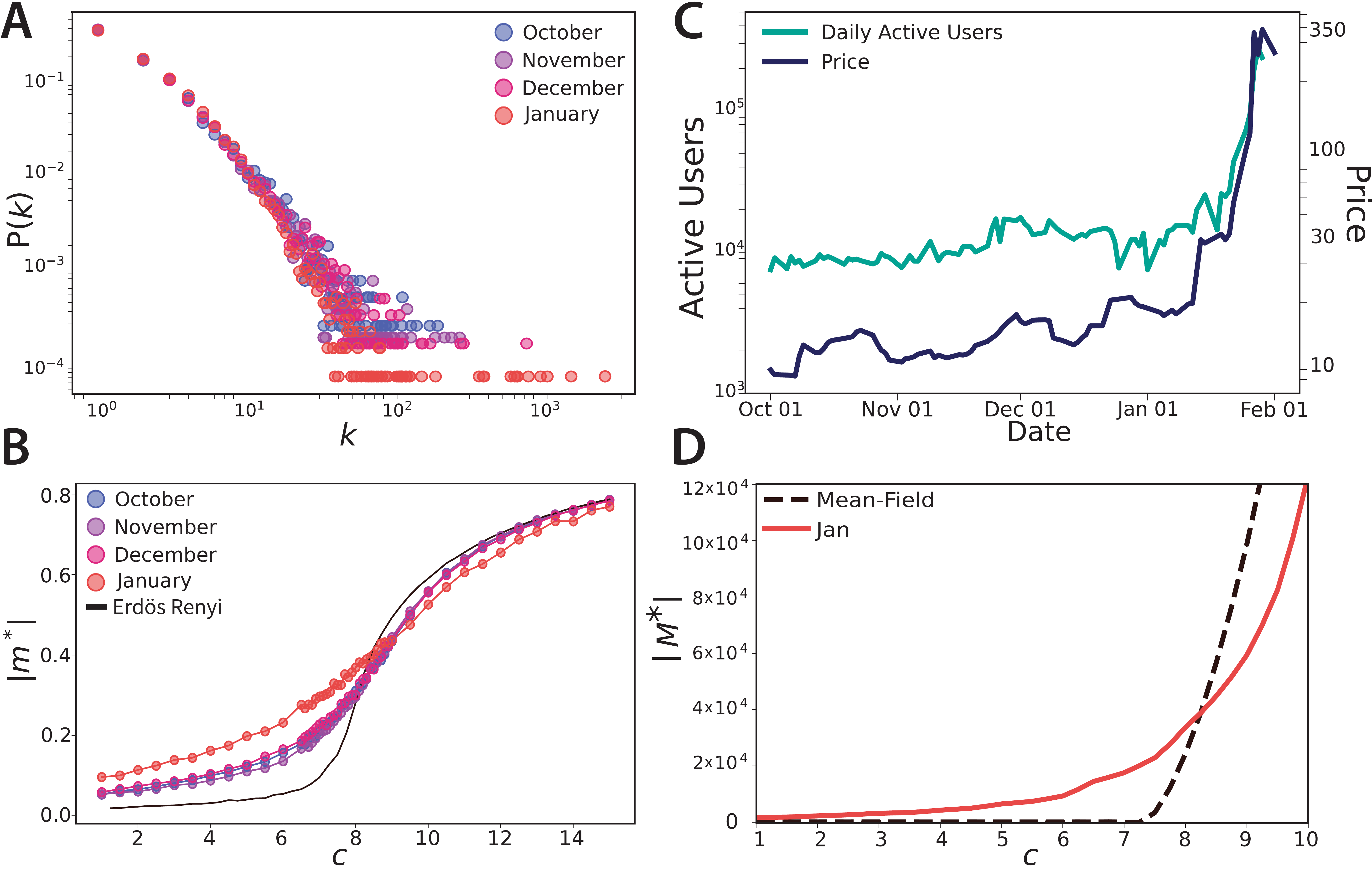}
    \caption{\textbf{WSB user interaction network and emergence of consensus.} 
A) Degree distributions of the monthly user interaction ('reply-to') networks, statistically validated using the disparity filter. B) Phase transition of the magnetization, obtained by simulating the model (for $\lambda=0.1$) on the monthly networks, as compared to the transition observed on \ER graphs. C) Number of daily active users (who contribute at least one post/comment) and daily closing price of GME shares. D) Phase transition of the extensive order parameter of the model (total magnetization for a community that grows as an exponential of $m$) according to the mean-field solution and to numerical simulations on the January user network.}
    \label{fig4}
\end{figure}

\subsection*{Consensus on the WSB user network.} 
We finally study how the model behaves on user-user interaction networks extracted from WSB conversation data. We build a network for each month by placing a directed link between two users $i$ and $j$ weighted by the number of times $i$ commented on $j$'s posts/comments during that period (Figure \ref{fig1}C). We then extract the most significant connections using the \emph{disparity filter} \cite{serrano2009extracting} with significance level $\alpha = 0.1$ (see Methods). 
We focus on the four months preceding the GME short squeeze, from October 2020 to January 2021.
Due to the explosion of activity in WSB, the network in January has many more nodes than those in the previous months; however, the application of the disparity filter makes their density of connections comparable (see Table \ref{tab:monthly_graph}).
In particular, all four networks display a power law distribution of the connectivity (Figure \ref{fig4}A), with some deviations caused by many super-hubs appearing in January.
Model simulations on these networks reported in Figure \ref{fig4}B (see the Supplementary Note 5 for further details) show that the degree heterogeneity of real user interactions leads to the emergence a non-negligible level of consensus also for very small values of user engagement $c$. 
This vanishing of the transition point is reminiscent of what occurs for other processes on scale-free networks, such as epidemic spreading \cite{pastorsatorras2001epidemic} and coordination games \cite{cimini2017evolutionary}.
Notably, the curves for the networks of October, November and December collapse onto each other, as opposed to that of January which is smoother due to the presence of more super-hubs and of a disassortative pattern \cite{newman2003miximg} (see Table \ref{tab:monthly_graph}) that ease the formation of an initial consensus. 
However magnetization alone does not allow for an appropriate comparison between networks of different sizes, since it represents the average opinion and is therefore an intensive variable. 
Equally important is the extent of consensus in terms of number of users. 
Indeed the success of the short squeeze required a large number of investors who bought and held GME shares. Another factor to take into account is the steep growth of the number of WSB users in correspondence with the short squeeze, as shown in Figure \ref{fig4}C. 
Altogether these observations suggest to consider an extensive order parameter, namely the total sum of opinions
within a population that grows with the level of consensus reached: $M^*=m^*\,N_0 e^{q|m^*|}$ (see Methods). 
As shown in Figure \ref{fig4}D this extensive magnetization features an abrupt transition, properly describing a sudden and large-scale formation of consensus. For a user engagement level $c$ that grows linearly in time (see Figure \ref{fig2}C), this transition is qualitatively similar to the sharp surge of GME price (see Figure \ref{fig4}C),  which ultimately represents the best proxy for the success of the short squeeze.

\section*{Discussion}

We remark that, in line with the standard statistical physics approach to social dynamics \cite{castellano2009statistical}, the proposed Voter-like framework is not intended as a falsifiable model of the real microscopic dynamics of the short squeeze on WSB. Rather it represents a minimal model that offers a possible explanation for key statistical patterns of the collective coordination action. In particular the model is intended to show that if the opinion dynamics is based on peer imitation, and if engagement or sentiment (corresponding to the control parameter $c$) grows up to the point of activating a self-induced global field, then it is possible to obtain an abrupt formation of consensus. The resulting order-to-disorder transition (depending on whether the global field activates or just represents noise) cannot be reproduced with traditional Voter frameworks, where consensus is reached very slowly due to a diffusive dynamics. Unfortunately, empirical validation of this model is impossible using solely WSB conversation data, for two reasons: i) we do not know how to map engagement/sentiment to the control parameter, and ii) since the user opinion (the order parameter) corresponds to participation or not to the short squeeze, we would need to know users' purchase transaction data, which are obviously not available due to privacy reasons.

The empirical evidence we are able to provide with our data is shown in the Results section above as well as in Supplementary Note 6. Figure \ref{fig3}C shows that the control parameter, as represented by the mean sentiment, grows as we approach the short squeeze date. Supplementary Figure 6 supports the peer imitation dynamics at the basis of the Voter model, using the sentiment of conversations among interacting users (rather than their opinion, for which we have no data). Similar evidence of social influence on the sentiment of comments by WSB users has been also reported in \cite{semenova2021reddits}. Figure \ref{fig4}D shows that the phase transition of the extensive magnetization is qualitatively comparable to the only empirical data available to proxy the real dynamics of users opinions: the price hike of GME shares. At last, in Supplementary Figure 7 we further assume to know a linear mapping between the mean sentiment and the control parameter $c$ to perform a more quantitative comparison. This latter exercise follows the same setup of the various works \cite{lyocsa2021YOLO,anand2021wallstreet,betzer2021if,hu2021rise,long2021just,wang2021predicting,semenova2021reddits} that try to use sentiment to predict stock price movements, however with a different aim: showing that the abrupt formation of consensus in the model is qualitatively similar to the GME price change due to the short squeeze.

\section*{Conclusion}

The empirical and theoretical results presented in this work can be useful to better understand the dynamics of consensus formation and collective actions on social networks. These phenomena have become increasingly relevant in recent years and have entered the financial domain with the GME case. This event is unlikely to remain isolated, particularly in the current financial context which sees the growing influence of retail and non-professional investors due to the emergence of commission-free trading and leverage platforms. While the ethical aspects of the ``democratization of trading and investing'' \cite{aramonte2021rising} and the ``David vs Goliath'' contrast between small investors versus hedge funds \cite{pelaez2021David} can be widely debated, their effects on market quality are certainly tangible \cite{eaton2021retail,allen2021squeezing}.

An inherent limitation of our empirical analyses is that we have focused on a single unprecedented financial mass action. Although we have briefly shown some similar case studies as well as counterexamples, a more in-depth analysis of several (possibly future) events of the same type can help corroborate or falsify our findings \cite{lamorgia2021doge}. 
Another empirical limitation is that we have derived the users interaction network from Reddit conversation data, while part of the coordination of the GME operation happened on other platforms as well, like Discord.
From the theoretical viewpoint, besides the model validation issues discussed above, it would be interesting to consider other popular models of opinion dynamics, such as the majority-vote model \cite{oliveira1992isotropic} or the threshold model \cite{granovetter1978threshold}, with the addition of a self-induced global field. 
Additionally, more realistic models can be developed that couple the dynamics of opinion formation with a growing size of the population.
At last, there is the very practical question of understanding how the dynamics of mass coordination reflect quantitatively on financial market movements. All these issues certainly represent interesting directions for future research \cite{gianstefani2022echo,sornette2022nonnormal}.

\section*{Methods}

\subsection*{Dataset.} 
We retrieved Reddit conversation data from Pushshift \cite{baumgartner2020pushshift}, an API that regularly copies activity data of Reddit and other social networks. We queried the service to retrieve information about WSB posts and comments (summarized in Supplementary Table 1) from September 01, 2019 to February 01, 2021. Note that our dataset covers the days following the short squeeze (25, 26, 27 January 2021) which were released by the service only on August 31, 2021. Overall our data contains 22\,099\,235 comments and 865\,597 posts. The dataset was cleaned by removing posts/comments by Reddit bots (Supplementary Table 2) as well as by "[deleted]" users (\ie, users who deleted their account before Pushshift could acquire their contributions). This latter operation was performed only for the analyses that required a unique userID (\ie, user activity statistics and user-user interaction networks), but not for those analyses that considered each post or comment on its own (\ie, tree statistics, ticker occurrences and sentiment). Data on stock price and traded volumes (Supplementary Table 3) for GME and other tickers were retrieved from the API service of \url{https://polygon.io}.

\subsection*{Ticker occurrences and Z-scores.}
To measure the popularity of a given stock in WSB conversations we computed $x_s(t)$, the count of how many times the ticker symbol of the stock $s$ (\eg, 'GME' for GameStop) appears as a regular expression in the raw text of posts/comments of day $t$. The mean $\mu_s(t) = \frac{1}{t} \sum_{t' = 1}^{t} x_s(t')$ and variance $\sigma^2_s(t)=\frac{1}{t} \sum_{t' = 1}^{t} \left[ x_s(t') - \mu_s(t) \right]^2$ of the time series $x_s(t)$ (starting from March 01, 2020) are used to obtain the Z-score $Z_s(t) = [x_s(t) - \mu_s (t)]/\sigma_s(t)$. The baseline $\overline{Z}(t)$ is the average Z-score of all stocks on day $t$ (computed over tickers with a symbol of at least three characters and appearing more than 10 times over the whole time interval).

\subsection*{Sentiment analysis and VADER lexicon.} VADER (\emph{Valence Aware Dictionary and sEntiment Reasoner}) \cite{hutto2014VADER} is an algorithm that assigns a piece of text with a compound score between $-1$ (very negative) and $+1$ (very positive). 
VADER is sensitive to both the polarity and intensity of the text, taking into account punctuation and word shape (ALL CAPS) used to add emphasis, degree modifiers that alter intensity (boosters such as "very" and dampeners such as "kind of"), slang and acronyms. 
VADER is based on a lexicon of words and emojis, each with an associated score ranging from -4 to +4 according to its meaning (from negative to positive). 
In line with other studies \cite{wang2021predicting,anand2021wallstreet}, we adapted VADER to the typical jargon and sarcasm of WSB users by adding to its lexicon the words reported in Supplementary Table 4. 
In Supplementary Note 3 we show the importance of considering such modified lexicon to detect the patterns of growing sentiment, also excluding the presence of biases arising from changes of jargon during time.

\subsection*{Voter model with self-induced global field.} 

The analytic mean-field solution of the model (full calculations in the Supplementary Note 7) leads to the following stochastic differential equation for the evolution of the magnetization of the system:
\begin{equation}
dm = v(m)dt + \sqrt{D(m)}dW
\end{equation}
where $W$ is the standard Wiener process while the drift and diffusion coefficients are 
\begin{equation}
\begin{split}
v(m) &= \lambda [f_c(m) - m ]\\
D(m) &= \frac{1}{N} \left\{ (1-\lambda)(1-m^2) + \lambda \left[ 1 - mf_c(m) \right] \right\}
\end{split}
\end{equation}
with $f_c(m) = 2P_1(e) -1 = \frac{c^m - 1}{c^m + 1}$. The formal solution for $m\simeq 0$ is
\begin{equation}
m_{t}=\frac{e^{-\lambda(1-\ln\sqrt{c})t}}{\sqrt {2N\lambda(1-\ln\sqrt{c})}}W_{e^{2\lambda(1-\ln\sqrt{c})t}-1}
\end{equation}
For $c \leq e^2$ the sign of the drift coefficient is always opposite to the sign of $m$ and the only equilibrium point is $m = 0$. For $c \geq e^2$ zero is no longer a stable point, for as soon as $m \neq 0$ the drift pushes the system towards a new equilibrium. $\lambda$ sets the speed at which the new stationary state is reached: the larger $\lambda$ the stronger the drift so the quicker the system will reach equilibrium. The critical value $c=e^2$ corresponds to a purely diffusive process driven by $D(m)$, which is of order $O\left(\frac{1}{N}\right)$ and thus can always be neglected except at the critical point. $D(m)$ is what drives the system out of the initial equilibrium state $m = 0$, but as soon as $m \neq 0$ the drift kicks in, governing the evolution of $m$.

Model simulations always start from an initial disordered configuration where each user is randomly assigned opinion $s=\pm 1$ with equal probability, such that $m(0)\simeq 0$. The dynamics is run until the magnetization reaches the stationary value $m^*$. Values shown in plots are averaged over 1000 independent runs. The extensive magnetization $M^*=m^*\, N$ is defined using a population size $N$ that grows exponentially with $|m^*|$ from a baseline level $N_0$. Parameters used for $N=N_0 e^{q|m^*|}$ are $N_0=10\,000$ (roughly the number of active users before January, see Figure \ref{fig3}C) and $q=6$, in order to have $N\approx 200\,000$ when $|m^*|\simeq 0.5$.
We remark that this change of $N$ is modeled ex-post, as in simulations the number of users is fixed. A possibly more realistic scenario would consists in having new nodes entering the system as the voter dynamics takes place. However it is reasonable to expect that a growing population may have an effect on the reaching of consensus only when the system is close to the critical point, \ie, in the region where the drift vanishes.

\subsection*{User network construction.}
For each month we reconstruct the network of social interactions by considering only posts and comments contributed during that month. Each user who contributed at least one of these posts/comments is represented as a node; each weighted directed link $w_{ij}$ represents the number of times user $i$ commented on posts/comments by user $j$. In order to filter out the less informative links and keep only those that are more likely to represent a significant interaction, we firstly removed all users who commented just once and then extracted the network backbone through the \emph{disparity filter} \cite{serrano2009extracting}. This algorithm assesses the statistical significance of links with respect to a null model where the weights of the links originating from a node are produced by a random assignment from a uniform distribution. Specifically, a link is deemed statistically significant if it satisfies $\alpha_{ij} = 1 - (k_i -1) \int_{0}^{w_{ij}/s_i} (1-x)^{k_i-2} dx < \alpha$, where $\alpha$ is the significance level and $k_i$ and $s_i = \sum_{j} w_{ij}$ are respectively the degree and strength of node $i$. 
The statistically validated network is then a binary undirected network made up of those links which at least in one direction satisfy the former condition (in the case where a node $i$ with $k_i=1$ is connected to a node $j$ with $k_j > 1$, the link is kept only if node $j$ satisfies the criterion). Note that in the case of a directed network the incoming and outgoing links associated with a node must be considered separately. At last, in our simulations we considered only the largest connected component of the network (see Table \ref{tab:monthly_graph}).

Deriving the interaction network of users from direct reply-to information is a popular approach in the literature on Twitter, where interactions inferred in this way can be validated against the follower/following relationships among users (when these are available) \cite{sousa2010characterization}. On top of this approach we also use statistical validation to filter out less-frequent interactions, according to user activity \cite{becatti2019extracting}. 
We remark that other approaches to infer user interactions have been used in the literature, such as connecting users who belong to the same comment chain and whose distance is within two or three comments of each other \cite{hamilton2017loyality}. We have checked that using this approach leads to results that are quantitatively very close to those reported in Figure \ref{fig4}B. 
Indeed the topological details of the user interaction network only have a second-order effect on the shape of the magnetization transition, as the phenomenon is qualitatively the same also in random \ER networks.

\begin{table}[h!]
	\centering
	\begin{tabular}{l|rr|rr|rrcccc}
		\toprule
		\textbf{Month} & $N_{tot}$ & $E_{tot}$ & $N_{b}$ & $E_{b}$ & $N_{g}$ & $E_{g}$ & $\langle k \rangle$ & $\langle k^2 \rangle/\langle k \rangle$ & $r$ & $\gamma$ \\
		\midrule
		October & 35850 & 564268 & 4235 & 11212 & 3542 & 8524 & 4.8 & 23.9 & 0.01 & 2.5 \\
		November & 47536 & 761374 & 5765 & 14888 & 4730 & 11307 & 4.7 & 27.5 & -0.03 & 2.4 \\
		December & 57822 & 913388 & 6675 & 18057 & 5474 & 13689 & 4.9 & 45.6 & -0.05 & 2.4 \\
		January & 357039 & 3359500 & 17740 & 38365 & 12232 & 29504 & 4.7 & 231.1 & -0.15 & 2.6 \\
		\hline 
	\end{tabular}
	\caption{Size of the monthly user-user interaction networks. The subscript \emph{tot} stands for the unfiltered data, $b$ for the backbone extracted with the disparity filter, $g$ for the largest connected component of the graph. Network statistics reported for these latter networks are the average degree $\langle k \rangle$, the average excess degree $\langle k^2 \rangle/\langle k \rangle$, the assortativity coefficient $r$ and the slope of the degree distribution $\gamma\simeq -\ln P(k)/\ln k$ (fitted starting from $k_{min}=10$).}
	\label{tab:monthly_graph}
\end{table}

\section*{Author contributions statement}

A.M. and A.D. performed the analysis and realised the figures. A.D. gather the data. R.D.C and G.C. designed the analysis and supervised the project.  G.C., R.D.C., A.M., A.D. wrote the paper. All authors discussed the results and contributed to the final manuscript.

\section*{Competing interests} The authors declare no competing interests. 

\section*{Data availability}
Reddit conversation data used in this study can be retrieved from the Pushshift API at \url{https://www.reddit.com/r/pushshift/}. Stock price and traded volumes data  are instead obtained by the Polygon API at \url{https://polygon.io}.

\section*{Acknowledgments}
This work has been supported by the ``Deep 'N Rec''  Progetto di Ricerca di Ateneo, University of Rome Tor Vergata. The icons in figure 1A-B-C and 3A have been designed by Freepik.

\end{document}